\documentclass[twocolumn,aps,prl,amssymb,showpacs]{revtex4}

\usepackage{graphicx}%
\usepackage{dcolumn}
\usepackage{amsmath}

\begin{document}

\title{Relevance of electron-lattice coupling in cuprates superconductors}
\author{T. Schneider$^{\text{1}}$, R. Khasanov$^{\text{1,2}}$, K. Conder$^{\text{3}}$%
, and H. Keller$^{\text{1}}$}
\address{$^{\text{(1)}}$ Physik-Institut der Universit\"{a}t Z\"{u}rich,
Winterthurerstrasse 190, CH-8057,Switzerland\\ $^{\text{(2)}}$
Paul Scherrer Institut, Labor f\"{u}r Myon-Spin Spektroskopie,
CH-5232 Villigen PSI, Switzerland\\$^{\text{(3)}}$ Laboratory for
Neutron Scattering, ETH Z\"{u}rich and PSI Villigen, CH-5232
Villigen PSI, Switzerland}

\begin{abstract}
We study the oxygen isotope ($^{16}$O,$^{18}$O) and finite size
effects in Y$_{1-x}\Pr_{x}$Ba$_{2}$Cu$_{3}$O$_{7-\delta }$ by
in-plane penetration depth ($\lambda _{ab}$) measurements. A
significant change of the length $L_{c}$ of the superconducting
domains along the $c$-axis and $\lambda _{ab}^{2}$ is deduced,
yielding the relative isotope shift $\Delta L_{c}/L_{c}\approx
\Delta \lambda _{ab}^{2}/\lambda _{ab}^{2}\approx -0.14$ for
$x=0,\ 0.2$ and $0.3$. This uncovers the existence and relevance
of the coupling between the superfluid, lattice distortions and
anharmonic phonons which involve the oxygen lattice degrees of
freedom.
\bigskip
\end{abstract}


\maketitle

 Since the discovery of superconductivity in cuprates by Bednorz and
M\"{u}ller\cite{bed} a tremendous amount of work has been devoted
to their characterization. The issue of inhomogeneity and its
characterization is essential for several reasons, including:
First, if inhomogeneity is an intrinsic property, a
re-interpretation of experiments, measuring an average of the
electronic properties, is unavoidable. Second, inhomogeneity may
point to a microscopic phase separation, i.e. superconducting
domains, embedded in a nonsuperconducting matrix. Third, there is
neutron spectroscopic evidence for nanoscale cluster formation and
percolative superconductivity in various
cuprates\cite{mesot,furrer}. Fourth, nanoscale spatial variations
in the electronic characteristics have been observed in underdoped
Bi$_{2}$Sr$_{2}$CaCu$_{2}$O$_{8+\delta }$ with scanning tunnelling
microscopy (STM)\cite{liu,chang,cren,lang}. They reveal a spatial
segregation of the electronic structure into 3nm diameter
superconducting domains in an electronically distinct background.
On the contrary, a large degree of homogeneity has been observed
by Renner and Fischer\cite{renner}. As STM is a surface probe the
relevance of these observations for bulk and
thermodynamic properties remains to be clarified. Fifth, in YBa$_{2}$Cu$_{3}$%
O$_{7-\delta }$, MgB$_{2}$, 2H-NbSe$_{2}$ and Nb$_{77}$Zr$_{23}$
considerably larger domains have been established. The magnetic
field
induced finite size effect revealed lower bounds ranging from $L=182$ to $%
814 $A \cite{book,tsfs}. Sixth, since the change of the lattice
parameters upon oxygen isotope exchange is negligibly
small\cite{conderla,raffa}, the occurrence of a significant
modification of the domains spatial extent, will provide clear
evidence for superconductivity mediated by local lattice
distortions.

This letter concentrates on
Y$_{1-x}\Pr_{x}$Ba$_{2}$Cu$_{3}$O$_{7-\delta }$ with $^{16}$O and
$^{18}$O. It addresses these issues by providing clear evidence
for a finite size effect on the in-plane London penetration depth,
revealing the existence of superconducting domains with spatial
nanoscale extent and its significant change upon oxygen isotope
exchange.

\begin{figure}
\centering
\includegraphics[width=0.95\linewidth]{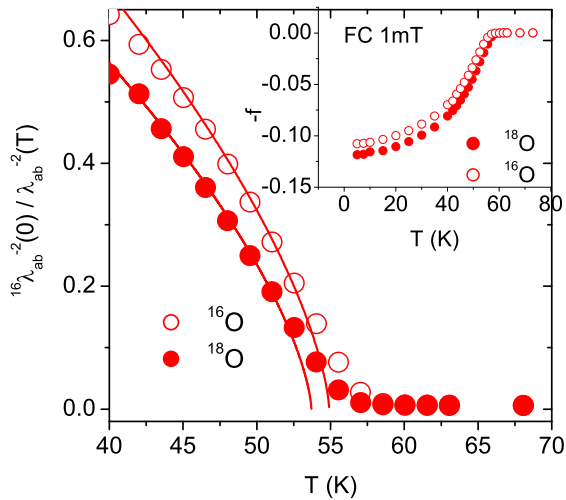}
\caption{$\left( ^{16}\lambda _{ab}(0)/\lambda _{ab}(T)\right)
^{2}$ versus $T$ for the  $^{16}$O and $^{18}$O samples of
Y$_{0.7}$Pr$_{0.3}$Ba$_{2}$Cu$_{3}$O$_{7-\delta }$. The solid
lines indicate the leading critical behavior of the homogeneous
systems as explained in the text. Insert: Meissner fraction $f$
versus temperature (FC 1mT). The error bars are smaller than the
size of the data points.} \label{fig1}
\end{figure}

Polycrystalline samples of
Y$_{1-x}$Pr$_{x}$Ba$_{2}$Cu$_{3}$O$_{7-\delta }$~ ($x=0$, 0.2,
0.3) were synthesized by solid-state reactions \cite{conder} and
their phase-purity was examined using powder x-ray diffraction.
For each doping concentration the samples were reground in a
mortar for about 60min. Powder samples with a grain size of
$<10\mu $m were obtained by using a system of 20/15/10$\mu $m
sieves. Oxygen isotope exchange was performed by heating the
powder in $\ ^{18}$O$_{2}$ gas. In order to ensure the same
thermal history of the substituted ($^{18}$O) and not substituted
($^{16}$O) samples, two experiments (in $^{16}$O$_{2}$ and
$^{18}$O$_{2}$) were always performed simultaneously
\cite{conder}. To achieve complete oxidation the exchange
processes were carried out at 550$^{o}$C during 30~h, followed by
slow cooling (20$^{o}$C/h). The $^{18}$O~content, determined from
a change of the sample weight after the isotope exchange, was
found to be 89(2)\% for all samples. Field-cooled (FC)
magnetization measurements were performed with a Quantum Design
SQUID magnetometer in field range 0.5 to 10mT and a temperature
range 5K to 100K. The powder samples ($\sim $100mg) were put in a
quartz ampule. To guarantee the same experimental conditions
(sample geometry and the background signal from the sample
holder), the same ampule was used. The absence of weak links
between grains was confirmed by the linear scaling of the FC
magnetization measured at 5K in 0.5mT, 1mT and 1.5mT. The Meissner
fraction $f$ \ was calculated from the mass and the x-ray density,
assuming spherical grains. An example of $f$ versus $T$ is
displayed in the insert of Fig.\ref{fig1}. Assuming spherical
grains of radius $R$, the data were analyzed on the basis of the
Shoenberg formula \cite{shoenberg}, allowing to calculate the
temperature dependence of the effective penetration depth $\lambda
_{eff}\left( T\right) /\lambda _{eff}\left( 0\right) $. For
sufficiently anisotropic extreme type II superconductors,
including Y$_{1-x}$Pr$_{x}$Ba$_{2}$Cu$_{3}$O$_{7-\delta }$ ,
$\lambda _{eff}$ is proportional to the in-plane penetration
depth, so that $\lambda _{eff}=1.31\lambda
_{ab}$\cite{Bafford88,Fesenko91}. Since it is not possible to
extract the absolute value of $\lambda _{ab}$ from our
measurements we take $\lambda _{ab}\left( 0K\right) $ from $\mu
$SR
measurements\cite{khasanov} to normalize the data. In the main panel of Fig.%
\ref{fig1} we displayed the resulting temperature dependence of $%
^{16}\lambda _{ab}^{2}\left( 0\right) /\lambda _{ab}^{2}\left(
T\right) $
for the $^{16}$O and $^{18}$O samples of Y$_{0.7}$Pr$_{0.3}$Ba$_{2}$Cu$_{3}$O%
$_{7-\delta }$. For comparison we included $^{16}\lambda
_{ab}^{2}\left( 0\right) /\lambda _{ab}^{2}\left( T\right) =$
$^{16}\lambda _{ab}^{2}\left( 0\right) /\lambda _{0,ab}^{2}\left(
1-T/T_{c}\right) ^{\nu }$ with the critical exponent $\nu =2/3$,
$^{16}T_{c}=54.9K$ and $^{18}T_{c}=53.7K$ to indicate the
asymptotic critical behavior for an infinite superconducting
domain. Apparently, the data are inconsistent with such a sharp
transition. It clearly uncovers a rounded phase transition which
occurs smoothly and with that a finite size effect at work. In
this context it is important to
emphasize that this finite size effect is not an artefact of Y$_{1-x}$Pr$%
_{x} $Ba$_{2}$Cu$_{3}$O$_{7-\delta }$ or the particular technique
used to evaluate $1/\lambda _{ab}^{2}\left( T\right) $. Indeed,
this rounding is
also seen in the data for\ YBa$_{2}$Cu$_{3}$O$_{7}$ \cite{panagop}, La$%
_{2-x} $Sr$_{x}$CuO$_{4}$ with x=0.1, 0.15 and 0.2\cite{pana}, and Bi$_{2}$Sr%
$_{2}$CaCu$_{2}$O$_{8+\delta }$ single
crystals\cite{jacobs,osborn}. Moreover, independent evidence for
superconducting domains of finite extent, stems from the analysis
of specific heat data\cite{tsfs}. Given the mounting evidence for
these domains, their behavior upon oxygen isotope exchange is
expected to offer valuable clues on the relevance of local lattice
distortions in the mechanism mediating superconductivity.

To elucidate this issue we perform a finite size scaling analysis
of the
in-plane penetration depth data for Y$_{1-x}$Pr$_{x}$Ba$_{2}$Cu$_{3}$O$%
_{7-\delta }$ with $^{16}$O and $^{18}$O. Supposing that cuprate
superconductors are granular, consisting of spatial
superconducting domains, embedded in a non-superconducting matrix
and with spatial extent $L_{a}$, $L_{b}$ and $L_{c}$ along the
crystallographic $a$, $b$ and $c$-axes, the correlation length
$\xi _{i}$ in direction $i$, increasing strongly when $T_{c}$ is
approached cannot grow beyond $L_{i}$. Consequently, for finite
superconducting domains, the thermodynamic quantities like the
specific heat and penetration depth are smooth functions of
temperature. As a remnant of the singularity at $T_{c}$ these
quantities exhibit a so called finite size effect\cite{fisher},
namely a maximum or an inflection point at $T_{p_{i}}$, where $\xi
_{i}\left( T_{p_{i}}\right) =L_{i}$. There is mounting
experimental evidence that for the accessible temperature ranges,
the effective finite temperature critical behavior of the cuprates
is controlled by the critical point of uncharged superfluids
(3D-XY)\cite{book,osborn}. In this case there is the universal
relationship
\begin{equation}
\frac{1}{\lambda _{i}^{2}\left( T\right) }=\frac{16\pi
^{3}k_{B}T}{\Phi _{0}^{2}\xi _{i}^{t}\left( T\right) },
\label{eq1}
\end{equation}
between the London penetration depth $\lambda _{i}$ and the
transverse correlation length $\xi _{i}^{t}$ in direction
$i$\cite{book,hohenberg}. As aforementioned, when the
superconductor is inhomogeneous, consisting of superconducting
domains with length scales $L_{i}$, embedded in a
non-superconducting matrix,\ the $\xi _{i}^{t}$'s do not diverge
but are bounded by
\begin{equation}
\xi _{i}^{t}\xi _{j}^{t}\leq L_{k}^{2},\ i\neq j\neq k.
\label{eq2}
\end{equation}
A characteristic feature of the resulting finite size effect is
the occurrence of an inflection point at $T_{p_{k}}$ below
$T_{c}$, the transition temperature of the homogeneous system.
Here
\begin{equation}
\xi _{i}^{t}\left( T_{p_{k}}\right) \xi _{j}^{t}\left(
T_{p_{k}}\right) =L_{k}^{2},\ i\neq j\neq k,  \label{eq3}
\end{equation}
and Eq.(\ref{eq1}) reduces to
\begin{equation}
\left. \frac{1}{\lambda _{i}\left( T\right) \lambda _{j}\left( T\right) }%
\right| _{T=T_{p_{k}}}=\frac{16\pi ^{3}k_{B}T_{p_{k}}}{\Phi _{0}^{2}}\frac{1%
}{L_{k}}.  \label{eq4}
\end{equation}
In the homogeneous case $1/\left( \lambda _{i}\left( T\right)
\lambda _{j}\left( T\right) \right) $ decreases continuously with
increasing temperature and vanishes at $T_{c}$, while for
superconducting domains, embedded in a non-superconducting matrix,
it does no vanish and exhibits an inflection point at
$T_{p_{k}}<T_{c}$, so that
\begin{equation}
\left. d\left( \frac{1}{\lambda _{i}\left( T\right) \lambda
_{j}\left( T\right) }\right) /dT\right|
_{T=T_{p_{k}}}=\text{extremum}  \label{eq5}
\end{equation}

\begin{figure}
\centering
\includegraphics[width=0.95\linewidth]{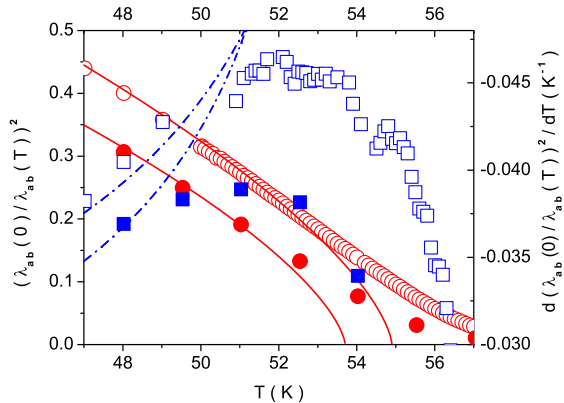} \vskip12pt
\caption{$^{16}\lambda _{ab}^{2}\left( T=0\right) /\lambda
_{ab}^{2}\left( T\right) $ $\left( \bigcirc :\text{ }^{16}\text{O,
}\bullet :\text{ }^{18}\text{O}\right) $ and $d\left( \lambda
_{ab}^{2}\left( T=0\right) /\lambda _{ab}^{2}\left( T\right)
\right) /dT$ $\left( \square :^{16}\text{O, }\blacksquare :\text{
}^{18}\text{O}\right) $versus $T$ for
Y$_{0.7}\Pr_{0.3}$Ba$_{2}$Cu$_{3}\ $O$_{7}$ with $^{16}$O and
$^{18}$O. The solid and dash dot lines indicate the leading
critical behavior of the homogeneous system as explained in the
text.}
\label{fig2}
\end{figure}

We are now prepared to perform the finite size scaling analysis of
the penetration depth data. In Fig.\ref{fig2} we displayed
$^{16}\lambda _{ab}^{2}\left( T=0\right) /\lambda _{ab}^{2}\left(
T\right) $ and $d\left( \lambda _{ab}^{2}\left( T=0\right)
/\lambda _{ab}^{2}\left( T\right) \right) /dT$ versus $T$ for
Y$_{0.7}$Pr$_{0.3}$Ba$_{2}$Cu$_{3}$O$_{7-\delta }$. The solid
lines are $\left( ^{16}\lambda _{ab}\left( 0\right) /^{16}\lambda
_{ab}\left( T\right) \right) ^{2}=1.62\left( 1-T/^{16}T_{c}\right)
^{\nu }$, $\left( ^{16}\lambda _{ab}\left( 0\right) /^{18}\lambda
_{ab}\left( T\right)
\right) ^{2}=1.4\left( 1-T/^{18}T_{c}\right) ^{\nu }$with $\ \nu =2/3$, $%
^{16}T_{c}=54.9$K, $^{18}T_{c}=53.7$K and \ the dash dot lines the
corresponding derivatives, indicating the leading critical
behavior of a domain, infinite in the c-direction. The extreme in
the first derivative around $T\approx 52.1$K and $51$K for
$^{16}$O and $^{18}$O respectively clearly reveal the existence of
an inflection point, characterizing the occurrence of a finite
size effect in $1/\lambda _{ab}^{2}$(Eq.(\ref{eq5})). Using
Eq.(\ref{eq4}) and the estimates for $T_{p_{c}}$, $^{16}\lambda
_{ab}^{2}\left( T_{p_{c}}\right) /^{16}\lambda _{ab}^{2}\left( 0\right) $,\ $%
^{18}\lambda _{ab}^{2}\left( T_{p_{c}}\right) /^{16}\lambda
_{ab}^{2}\left( 0\right) $ and $^{16}\lambda _{ab}\left( 0\right)
$ listed in Table I, we obtain $^{16}L_{c}=19.5\left( 8\right)
$\AA\ and $^{18}L_{c}=22.6\left( 9\right) $\AA\ for the spatial
extent of the superconducting domains along the c-axis. Note that
the rather broad peak around the inflection point reflects the
small value of $L_{c}$. Indeed, in Bi2212, where the same analysis
gives $L_{c}\approx 68$\AA , this peak was found to be
considerably sharper\cite{tsbifs}.

\begin{figure}
\centering
\includegraphics[width=0.95\linewidth]{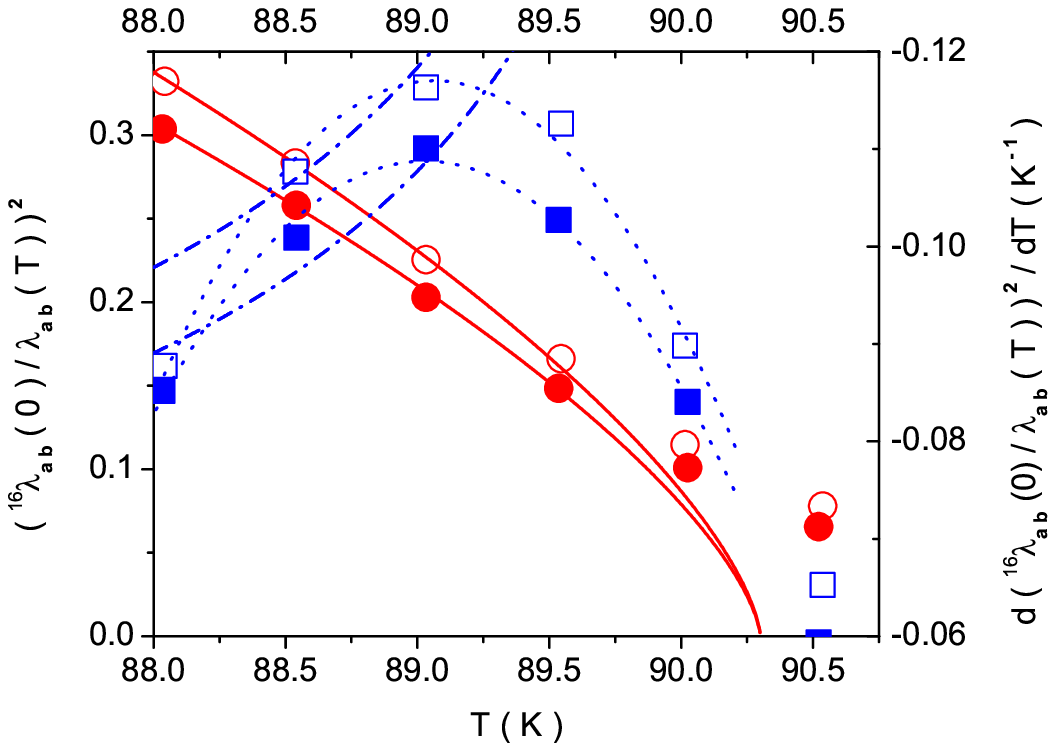} \vskip12pt
\caption{$^{16}\lambda _{ab}^{2}\left( T=0\right) /\lambda
_{ab}^{2}\left( T\right) $ $\left( \bigcirc :\text{ }^{16}\text{O,
}\bullet :\text{ }^{18}\text{O}\right) $ and $d\left( \lambda
_{ab}^{2}\left( T=0\right) /\lambda _{ab}^{2}\left( T\right)
\right) /dT$ $\left( \square :^{16}\text{O, }\blacksquare :\text{
}^{18}\text{O}\right) $ versus $T$ for YBa$_{2}$Cu$_{3}$O$_{7}$
with $^{16}$O and $^{18}$O. The solid and \ dash dot lines
indicate the leading critical behavior of the homogeneous system
as explained in the text. The dot lines are quadratic fits used to
determine the inflection points $T_{p_{c}}$.}
\label{fig3}
\end{figure}

To explore the dependence of this change on Pr concentration we
performed analogous magnetization measurements on
Y$_{1-x}$Pr$_{x}$Ba$_{2}$Cu$_{3}$O$_{7-\delta }$ with $x=0$ and
$x=0.2$ and extracted the in-plane penetration depth, as outlined
above. In Fig.\ref{fig3} we show $^{16}\lambda _{ab}^{2}\left(
T=0\right) /\lambda _{ab}^{2}\left( T\right) $ and $d\left(
\lambda _{ab}^{2}\left( T=0\right) /\lambda _{ab}^{2}\left(
T\right) \right) /dT$ versus $T$ for \
YBa$_{2}$Cu$_{3}$O$_{7-\delta }$ with $^{16}$O and $^{18}$O. The
solid lines are $\left( ^{16}\lambda _{ab}\left( 0\right)
/^{16}\lambda _{ab}\left( T\right) \right) ^{2}=3.9\left(
1-T/^{16}T_{c}\right) ^{2/3}$, $\left( ^{16}\lambda _{ab}\left(
0\right) /^{18}\lambda _{ab}\left( T\right) \right)
^{2}=3.55\left( 1-T/^{18}T_{c}\right) ^{2/3}$ with $\
^{16}T_{c}\approx ^{18}T_{c}=90.3$K and \ the dash dot lines the
corresponding derivatives, indicating the leading critical
behavior of the homogeneous system. The finite size estimates for
various quantities and the corresponding isotope shifts, for an
$^{18}$O content 89\% are summarized in Table I (we define the
relative oxygen isotope shift of a physical quantity $X$ as
$\Delta X / X = (^{18}X-^{16}X)/^{16}X$).  While the isotope
effect on $T_{c}$ and the inflection points $T_{p_{c}}$is very
small, there is an appreciable shift of $1/\lambda _{ab}^{2}$. As
indicated in Fig.\ref{fig3}, we used a quadratic fit around the
inflection
point to determine $T_{p_{c}}$. Nevertheless, the error left ($0.1$ to $0.2$%
K) leads to the main uncertainty in terms of $\lambda
_{ab}^{2}\left( T_{p_{c}}\right) /\lambda _{ab}^{2}\left( 0\right)
$. From Table I several
observations emerge. First, $L_{c}$ increases systematically with reduced $%
T_{p_{c}}$. Second, $L_{c}$ grows with increasing $x$ and upon
isotope exchange ($^{16}$O, $^{18}$O). Third, the relative shift
of $T_{p_{c}}$  is very small. This reflects the fact that the
change of $L_{c}$ is essentially
due to the superfluid, probed in terms of $\lambda _{ab}^{2}$. Accordingly, $%
\Delta L_{c}/L_{c}\approx \Delta \lambda _{ab}^{2}/\lambda _{ab}^{2}$ for $%
x=0,\ 0.2$ and $0.3$. Indeed the relative shifts of $T_{p_{c}}$,
$\lambda _{ab}^{2}\left( T_{p_{c}}\right) $ and $L_{c}$ are not
independent. Eq.(\ref {eq4}) implies,
\begin{equation}
\frac{\Delta L_{c}}{L_{c}}=\frac{\Delta T_{p_{c}}}{T_{p_{c}}}+%
\frac{\Delta \lambda _{ab}^{2}\left( T_{p_{c}}\right)}{\lambda
_{ab}^{2}\left( T_{p_{c}}\right) }.  \label{eq6}
\end{equation}

\begin{center}
\begin{tabular}{|c|c|c|c|}
\hline x & 0 & 0.2 & 0.3 \\ \hline $\Delta
T_{p_{c}}/T_{p_{c}}$ & -0.000(2) & -0.015(3) & -0.021(5) \\
\hline $\Delta L_{p_{c}}/L_{p_{c}}$ & 0.12(5) & 0.13(6) &
0.16(5) \\
\hline $\Delta \lambda _{ab}^{2}\left( T_{p_{c}}\right) /
\lambda_{ab}^{2}\left( T_{p_{c}}\right) $ & 0.11(5) & 0.15(6) &
0.15(5) \\
\hline $^{16}\lambda _{ab}^{2}\left( ^{16}T_{p_{c}}\right)
/^{16}\lambda _{ab}^{2}\left( 0\right) $ &
4.4(2) & 4.0(2) & 4.4(2) \\
\hline $^{18}\lambda _{ab}^{2}\left( ^{18}T_{p_{c}}\right)
/^{16}\lambda _{ab}^{2}\left( 0\right) $ &
4.9(2) & 4.6(2) & 5.2(2) \\
\hline $^{16}T_{p_{c}}(K)$ & 89.0(1) &
67.0(1) & 52.1(2) \\
\hline $^{18}T_{p_{c}}\left( K\right) $ &
89.0(1) & 66.0(2) & 51.0(2) \\
\hline
$^{16}L_{p_{c}}\left( {\rm \AA} \right) $ & 9.7(4) & 14.2(7) & 19.5(8) \\
\hline
$^{18}L_{p_{c}}\left(  {\rm \AA} \right) $ & 10.9(4) & 16.0(7) & 22.6(9) \\
\hline
$^{16}\lambda _{ab}\left( 0\right) \left(  {\rm \AA} \right) $ & 1250(10) & 1820(20) & 2310(30) \\
\hline
\end{tabular}
\end{center}

\bigskip

\bigskip

Table I: Finite size estimates for $^{16}T_{p_{c}}$, $^{16}T_{p_{c}}$, $%
\Delta T_{p_{c}}=^{16}T_{p_{c}}-^{18}T_{p_{c}}$,$\ ^{16}\lambda
_{ab}^{2}\left( ^{16}T_{p_{c}}\right) /^{16}\lambda
_{ab}^{2}\left( 0\right) $ and $\ ^{18}\lambda _{ab}^{2}\left(
^{18}T_{p_{c}}\right) /^{16}\lambda _{ab}^{2}\left( 0\right) $,
and the resulting relative shifts $\Delta
T_{p_{c}}/^{16}T_{p_{c}}$and $\Delta \lambda
_{ab}^{2}/^{16}\lambda _{ab}^{2}\left( ^{16}T_{p_{c}}\right) $ for
an $^{\text{18}}$O content of 89\%. $^{16}L_{p_{c}}$,
$^{18}L_{p_{c}}$ and $\Delta L_{p_{c}}/^{16}L_{p_{c}} $ follow
then via Eq.(\ref{eq4}). $^{16}\lambda _{ab}\left( 0\right) $ are
$\mu $SR estimates\cite{khasanov}

\bigskip
To appreciate the implications of these estimates, we note that
for fixed Pr concentration the lattice parameters remain
essentially unaffected\cite {conderla,raffa}. Accordingly, an
electronic mechanism, without coupling to local lattice
distortions and anharmonic phonons, implies $\Delta L_{c}=0$. On
the contrary, a significant change of $L_{p_{c}}$ upon oxygen
exchange uncovers the coupling to local lattice distortions and
anharmonic phonons involving the oxygen lattice degrees of
freedom. A glance to Table I shows that the relative change of the
domains along the c-axis upon oxygen isotope exchange is
significant, ranging from $12$ to $16$\%, while the relative
change of the inflection point or the transition temperature is an
order of magnitude smaller. For this reason the significant
relative change of $L_{c}$ at fixed Pr concentration is
accompanied by essentially the same relative change of $\lambda
_{ab}^{2}$, which probes the superfluid. This uncovers
unambiguously the existence and relevance of the coupling between
the superfluid, lattice distortions and anharmonic phonons
involving the oxygen lattice degrees of freedom. Potential
candidates are the Cu-O bond-stretching-type phonons showing
temperature dependence, which parallels that of the
superconductive order parameter\cite{chung}. Independent evidence
for the shrinkage of limiting length scales upon isotope exchange
stems from the behavior close to the quantum superconductor to
insulator transition where $T_{c}$ vanishes\cite{tsiso}. Here the
cuprates become essentially two dimensional and correspond to a
stack of independent slabs of thickness
$d_{s}$\cite{tseuro,tsphysicab}. It was found that the relative
shift $\Delta d_{s}/d_{s}$ upon isotope exchange adopts a rather
unique value, namely $\Delta d_{s}/d_{s}\approx
-0.03$\cite{tsiso}.

In conclusion, we reported the first observation of the combined
finite size and oxygen isotope exchange effects on the spatial
extent $L_{c}$ of the
superconducting domains along the c-axis and the in-plane penetration depth $%
\lambda _{ab}$. Although the majority opinion on the mechanism of
superconductivity in the cuprates is that it occurs via a purely
electronic mechanism involving spin excitations, and lattice
degrees of freedom are supposed to be irrelevant, we have shown
the relative isotope shift $\Delta L_{c}/^{16}L_{c}\approx \Delta
\lambda _{ab}^{2}/^{16}\lambda _{ab}^{2}\approx -0.15$ uncovers
clearly the existence and relevance of the coupling between the
superfluid, lattice distortions and anharmonic phonons which
involve the oxygen lattice degrees of freedom.

The authors are grateful to K.A. M\"{u}ller and J. Roos for very
useful comments and suggestions on the subject matter. This work
was partially supported by the Swiss National Science Foundation.
\\

\end{document}